\documentclass[preprint,superscriptaddress,showpacs,nofootinbibfloatfix,amsmath,amsfonts,amssymb]{revtex4-1}%
\usepackage{amsmath,amsfonts,amssymb,color}
\usepackage{amsthm}
\usepackage{leftidx}
\usepackage{graphicx}
\usepackage{xcolor}
\usepackage{dcolumn}
\usepackage{bm}
\usepackage{epstopdf}
\usepackage{epsfig}
\usepackage{environ}
\usepackage{pdfcomment}

\usepackage{multirow}
\usepackage{setspace}
\usepackage{color}

\usepackage{float}
\usepackage[T1]{fontenc}
\usepackage[latin9]{inputenc}
\usepackage{setspace}
\usepackage{esint}

\providecommand{\tabularnewline}{\\}

\begin{document}


\title{Non-Hermitian topological phases and dynamical quantum phase transitions: a generic connection}

\author{Longwen Zhou}
\email{zhoulw13@u.nus.edu}
\affiliation{%
	Department of Physics, College of Information Science and Engineering, Ocean University of China, Qingdao, China 266100
}
\author{Qianqian Du}
\affiliation{%
	Department of Physics, College of Information Science and Engineering, Ocean University of China, Qingdao, China 266100
}

\date{\today}

\begin{abstract}
The dynamical and topological properties of non-Hermitian systems
have attracted great attention in recent years. In this work, we establish
an intrinsic connection between two classes of intriguing phenomena
-- topological phases and dynamical quantum phase transitions (DQPTs)
-- in non-Hermitian systems. Focusing on one-dimensional models with
chiral symmetry, we find DQPTs following the quench from a trivial
to a non-Hermitian topological phase. Moreover, the
critical momenta and critical time of the DQPTs
are found to be directly related to the topological invariants of
the non-Hermitian system. We further demonstrate our theory in three
prototypical non-Hermitian lattice models, the lossy Kitaev chain
(LKC), the LKC with next-nearest-neighbor hoppings, and the nonreciprocal
Su-Schrieffer-Heeger model. Finally, we suggest a proposal to experimentally
verify the found connection by a nitrogen-vacancy center in
diamond.
\end{abstract}

\pacs{}
\keywords{}
\maketitle

\section{Introduction}\label{sec:Int}
Non-Hermitian systems have attracted great interest in recent years
due to their intriguing dynamical and topological properties~\cite{NHRev1,NHRev2,NHRev3,NHRev4,NHRev5,NHRev6,NHRev7,NHRev8}.
Theoretically, exceptional point~(EP) induced chiral dynamics~\cite{EPLoop1,EPLoop2,EPLoop3}
and non-Hermitian topological matter~\cite{NHTP1,NHTP2,NHTP3,NHTP4} have been
found and explored in a variety of systems. Experimentally, non-Hermitian
topological phases and phenomena have been observed in cold atom~\cite{NHCdAtm1,NHCdAtm2,NHCdAtm3},
photonic~\cite{NHPhoto1,NHPhoto2,NHPhoto3,NHPhoto4}, electric circuit~\cite{NHEC1,NHEC2,NHEC3}, 
acoustic~\cite{NHAcou1,NHAcou2,NHAcou3} systems and nitrogen-vacancy-center in diamond~\cite{NVExp0,NVExp1},
leading to potential applications like unidirectional transport devices~\cite{NHUniTrans1,NHUniTrans2}, 
topological lasers~\cite{TILZ1,TILZ2,TILZ3} and high-performance sensors~\cite{NHSens1,NHSens2,NHSens3,NHSens4}.

To date, non-Hermitian topological phases (NHTPs) have been classified
and characterized according to their protecting symmetries~\cite{NHTP2,NHTP3,NHTP4}.
Finding the dynamical signatures of these nonequilibrium topological
matter has become an urgent topic for further theoretical and experimental
explorations. In the literature, several dynamical probes to the topological
invariants of one- and two-dimensional non-Hermitian phases have been
proposed, such as the non-Hermitian extension of mean chiral displacements~\cite{ZhouMCD1,ZhouMCD2,ZhouMCD3} 
and dynamical winding numbers~\cite{DWN,ZhouDWN1,ZhouDWN2,ZhouDWN3}. In the meantime, DQPTs
(i.e., nonanalytic behaviors of certain observables in time domain~\cite{DQPTRev1,DQPTRev2,DQPTRev3,DQPTRev4})
following a quench across the EPs of a non-Hermitian lattice model
is investigated in Ref.~\cite{ZhouDQPT1}, and the monotonic growth of a dynamical
topological order parameter in time is observed if an isolated
EP is crossed during the quench~\cite{ZhouDQPT1}. This discovery indicates an
underlying relationship between the two notably different nonequilibrium
phenomena, NHTPs and DQPTs. However, the more general connection between
NHTPs and DQPTs, together with its possible experimental observations have
not been revealed.

In this work, we uncover an intrinsic connection between the topological
phases and DQPTs in one-dimensional (1D) non-Hermitian systems. In
Sec.~\ref{sec:Theory}, we develop our theoretical framework leading
to the establishment of this connection. In Sec.~\ref{sec:Model},
we demonstrate the found connection in three different non-Hermitian
lattice models, the lossy Kitaev chain (LKC), the LKC with next-nearest-neighbor
(NNN) hoppings and pairings, and the nonreciprocal Su-Schrieffer-Heeger
(NRSSH) model. In each model, a direct link between the bulk topological
invariant of a non-Hermitian phase and the number of critical time
and momenta of DQPTs following a quench to the corresponding phase is found. In
Sec.~\ref{sec:Exp}, we discuss an experimental setup, the nitrogen-vacancy
(NV) center in diamond, in which the discovered connection may
be tested. We conclude this work and discuss potential future directions
in Sec.~\ref{sec:Sum}.

\section{Theory}\label{sec:Theory}
In this section, we introduce a generic class of non-Hermitian lattice
models and describing the topological characterization of its bulk states in
Subsec.~\ref{subsec:NHTPs}.
In Subsec.~\ref{subsec:DQPTs}, we introduce relevant quantities to
characterize DQPTs in 1D non-Hermitian systems, and establish their connections with the underlying
topological properties of the system.

\subsection{NHTPs}\label{subsec:NHTPs}
We start with a non-Hermitian Hamiltonian $H\neq H^{\dagger}$, which
describes particles in a 1D lattice subjecting to gains, losses and/or
nonreciprocal effects. Under the periodic boundary condition, we can
express the Hamiltonian of the system as $H=\sum_{k}\Psi_{k}^{\dagger}H(k)\Psi_{k}$,
where $k\in[-\pi,\pi)$ is the quasimomentum, $\Psi_{k}^{\dagger}$
($\Psi_{k}$) is the two-component creation (annihilation) operator
in momentum representation, and the Bloch Hamiltonian $H(k)$ takes
the general form
\begin{equation}
H(k)=[h_{a}(k)-ig_{a}(k)]\sigma_{a}+[h_{b}(k)-ig_{b}(k)]\sigma_{b}.\label{eq:Hk}
\end{equation}
Here $h_{a,b}(k)$ and $g_{a,b}(k)$ are real-valued functions of
the quasimomentum $k$, $i$ denotes the imaginary unit, and $\sigma_{a,b}$
are any two of the three Pauli matrices $\sigma_{x}$, $\sigma_{y}$,
and $\sigma_{z}$, with $\{\sigma_{a},\sigma_{b}\}=0$ for $a\neq b$. We will also
denote the $2\times2$ identity matrix as $\sigma_{0}$.

The non-Hermiticity of $H$ implies that $H(k)\neq H^{\dagger}(k)$
at a generic quasimomentum $k$. The dispersion relation of non-Hermitian Bloch
Hamiltonian $H(k)$ is given by
\begin{alignat}{1}
E_{\pm}(k)= & \pm\sqrt{[h_{a}(k)-ig_{a}(k)]^{2}+[h_{b}(k)-ig_{b}(k)]^{2}}\nonumber\\
= & \pm E(k).\label{eq:pmEk}
\end{alignat}
It is clear that $\pm E(k)$ are in general complex numbers. The spectrum
of $H(k)$ becomes gapless at zero energy if $E(k)=0$. According
to Eq.~(\ref{eq:pmEk}), this is achieved when both the following conditions
are satisfied
\begin{alignat}{1}
h_{a}^{2}(k)+h_{b}^{2}(k)-g_{a}^{2}(k)-g_{b}^{2}(k)& = 0,\label{eq:Gaples1}\\
h_{a}(k)g_{a}(k)+h_{b}(k)g_{b}(k)& = 0.\label{eq:Gaples2}
\end{alignat}
By solving these equations, we could obtain the quasimomentum $k_{0}$
at which the spectrum gap closes, and find the boundaries separating
different gapped phases in the parameter space, which could also be
the boundaries among different bulk topological phases of the system.

To characterize the topological properties of the gapped phases of
$H(k)$~(i.e., $E(k)\neq0$ for all $k$), the usual recipe is
to identify the symmetries of the system and construct the relevant
topological invariants. From the commutation relation of Pauli matrices
$[\sigma_{a},\sigma_{b}]=2i\epsilon_{abc}\sigma_{c}$, it is clear
that the $H(k)$ in Eq.~(\ref{eq:Hk}) possesses the chiral (sublattice)
symmetry ${\cal S}=\sigma_{c}$ ($c\neq a,b$), in the sense that
${\cal S}^{2}=\sigma_{0}$ and ${\cal S}H(k){\cal S}=-H(k)$. The
bulk topological phases of a non-Hermitian Bloch Hamiltonian with
sublattice symmetry ${\cal S}$ can usually be characterized by a
winding number $w$, defined as
\begin{equation}
w=\int_{-\pi}^{\pi}\frac{dk}{2\pi}\partial_{k}\phi(k),\quad\phi(k)\equiv\arctan\left[\frac{h_{b}(k)-ig_{b}(k)}{h_{a}(k)-ig_{a}(k)}\right],\label{eq:w}
\end{equation}
which describes the accumulated change of winding angle $\phi(k)$
throughout the first Brillouin zone (BZ). Note that the value of $w$
is always real even though $\phi(k)$ is in general complex, as the imaginary
part of $\phi(k)$ has no winding in the first BZ (see Ref.~\cite{DWN}
for a proof). Furthermore, the $w$ as defined in Eq.~(\ref{eq:w}) can
take either integer or half-integer values, depending on the relative
locations between the EPs of $H(k)$ on the $h_{a}$-$h_{b}$ plane
and the trajectory of vector ${\bf h}(k)=[h_{a}(k),h_{b}(k)]$ versus
$k$. That is, if ${\bf h}(k)$ encircles an even (odd) number of
EPs, we would have $w\in\mathbb{Z}$~($w\in(2\mathbb{Z}+1)/2$)~\cite{DWN}. 
Within a gapped topological phase of $H(k)$, the value
of $w$ is a constant, whereas it takes a quantized (or half-quantized
jump) when a phase boundary determined by Eqs.~(\ref{eq:Gaples1})-(\ref{eq:Gaples2})
is crossed. Therefore, the invariant $w$ yields a characterization
for all the bulk non-Hermitian topological phases of $H(k)$. Experimentally,
the winding number $w$ can be obtained by measuring the mean chiral
displacements of wave packets~\cite{ZhouMCD1} or the dynamical winding numbers
of time-averaged spin textures \cite{DWN}.

\subsection{DQPTs and their relations to NHTPs}\label{subsec:DQPTs}

DQPTs are characterized by nonanalytic behaviors of system observables
as functions of time. They are usually found in the dynamics following
a quench across the equilibrium phase transition point of a quantum
many-body system (see Ref.~\cite{DQPTRev1,DQPTRev2,DQPTRev3,DQPTRev4}
for reviews). The central object in the description of DQPTs is the
return amplitude $G(t)=\langle\Psi|U(t)|\Psi\rangle$, where $|\Psi\rangle$
is the initial many-particle state (usually taken as the equilibrium
ground state of the system before the quench) and $U(t)$ is the evolution
operator of the system following a quantum quench (or some other nonequilibrium
protocols). Formally, $G(t)$ mimics the dynamical partition function
of the post-quench evolution. When $G(t)=0$ at a critical time $t_{c}$,
the initial state evolves into its orthogonal state. The rate function
of return probability $g(t)=-\lim_{N\rightarrow\infty}N^{-1}\ln|G(t)|^{2}$
($N$ is the number of degrees of freedom of the system) or its time
derivatives would then become nonanalytic at $t=t_{c}$, signifying
a DQPT. Accompanying theoretical discoveries~\cite{DQPT1,DQPT2,DQPT3}, DQPTs have
been observed in cold atoms~\cite{DQPTExp3,DQPTExp4,DQPTExp9,DQPTExp10}, trapped ions~\cite{DQPTExp1,DQPTExp2}, superconducting
qubits~\cite{DQPTExp5}, nanomechanical and photonic systems~\cite{DQPTExp6,DQPTExp7,DQPTExp8}.
Recent studies further extend DQPTs to periodically driven (Floquet)
systems~\cite{DQPTExp11,FDQPT1,FDQPT2,FDQPT3,FDQPT4,FDQPT5}, accompanied by an experimental realization in the NV center
setup~\cite{DQPTExp11}.

To relate DQPTs with topological phases in non-Hermitian systems,
we focus on a unique class of quench protocol, in which the system
is initialized with equal populations but no coherence on the two
bands of the non-Hermitian Bloch Hamiltonian $H(k)$ in Eq.~(\ref{eq:Hk}),
i.e., an infinite-temperature initial state $\prod_{k\in{\rm BZ}}\rho_{0}$,
with $\rho_{0}=\sigma_{0}/2$ being the single-particle density matrix.
The evolution of $\rho_{0}$ at time $t>0$ is governed by $H(k)$,
and the return amplitude $G(k,t)$, defined as the expectation value
of evolution operator $U(k,t)=e^{-iH(k)t}$ over the initial state
$\rho_{0}$ reads
\begin{equation}
G(k,t)={\rm Tr}[\rho_{0}U(k,t)]=\cos[E(k)t],\label{eq:Gkt}
\end{equation}
where Eqs.~(\ref{eq:Hk}) and (\ref{eq:pmEk}) have been used to reach
the second equality. When possible DQPTs happen, we would have $\cos[E(k)t]=0$,
leading to the critical times
\begin{equation}
t_{n}(k)=\left(n-\frac{1}{2}\right)\frac{\pi}{E(k)},\qquad n\in\mathbb{Z}.\label{eq:tn}
\end{equation}
This seemingly innocent expression yields rather different predictions
for Hermitian and non-Hermitian systems. In a Hermitian system, where
the dispersion relation $E(k)$ is always real and positive, we would have
a set of critical times $t_{n}(k)$ for each quasimomentum $k$. However, the
resulting non-analyticity in the rate function $g(t)$
is simply originated from the oscillatory dynamics of a single Bloch
state rather then an actual phase transition, which only happens in
thermodynamic limit ($N\rightarrow\infty$). On the other hand, when
$H(k)$ is non-Hermitian, we have $E(k)\in\mathbb{C}$ in general,
and real critical times $t_{n}(k)$ emerge only at the critical momenta
$k_{c}$ where $E(k_{c})\in\mathbb{R}$. According to Eq.~(\ref{eq:pmEk}),
this is equivalent to the fulfillment of the following two conditions:
\begin{alignat}{1}
h_{a}^{2}(k)+h_{b}^{2}(k)-g_{a}^{2}(k)-g_{b}^{2}(k)& > 0,\label{eq:kc1}\\
h_{a}(k)g_{a}(k)+h_{b}(k)g_{b}(k)& = 0,\label{eq:kc2}
\end{alignat}
which only yield solutions at isolated values of $k$. For a critical
momentum $k_{c}$ satisfying both the Eqs.~(\ref{eq:kc1}) and (\ref{eq:kc2}),
we would have $G(k_{c},t_{n})=0$ for all $n\in\mathbb{Z}$. In the
thermodynamic limit, the rate function of return probability for the
many-particle initial state $\prod_{k\in{\rm BZ}}\rho_{0}$ is given
by
\begin{equation}
g(t)=-\lim_{N\rightarrow\infty}\frac{1}{N}\ln|G(k,t)|^{2}=-\int_{{\rm BZ}}\frac{dk}{2\pi}\ln|G(k,t)|^{2},\label{eq:gt}
\end{equation}
which will have discontinuous first-order time derivatives at all $t_{n}(k_{c})$.
Note that by taking the limit $N\rightarrow\infty$, the distribution
of $t_{n}(k)$ on the complex time plane changes from isolated points
to a continuous line, whose crossings along the real-time axis correspond
to the critical time of genuine DQPTs in the sense of Fisher zeros~\cite{DQPTRev1}.

The connection between DQPTs and NHTPs in chiral-symmetric 1D systems
becomes transparent at this stage. First, we note that the Eq.~(\ref{eq:Gaples2}),
which determines the gapless quasimomenta $\{k_{0}\}$ is identical
to Eq.~(\ref{eq:kc2}). This implies that the critical momenta $\{k_{c}\}$
of DQPTs can only be a subset of $\{k_{0}\}$. Second, plugging $\{k_{0}\}$
into Eq.~(\ref{eq:Gaples1}), we obtain an expression for the boundaries
separating different NHTPs in the parameter space. Combining this
with Eq.~(\ref{eq:kc1}) further suggests that DQPTs can only be observed
in certain regimes that are distinguished from the others by the topological
phase boundaries of $H(k)$. Third, since the number of gapless quasimomenta
$k_{0}$ is closely related to the change of topological invariant
$w$ in Eq.~(\ref{eq:w}) across the corresponding topological phase
transition point, DQPTs with different numbers of $k_{c}$ are expected
to happen following quenches to different NHTPs. As each critical
momentum $k_{c}$ determines a unique period $T(k_{c})=\pi/E(k_{c})$
for the DQPTs, the number of critical period $T(k_{c})$ is determined
by the number of distinct critical momenta. The third point then suggests a
way to distinguish different NHTPs through the quantitative difference
of the critical time-periods of DQPTs therein.

Putting together, we have uncovered an intrinsic relation between
the topological phases and DQPTs in non-Hermitian systems, which not
only bridges the gap between these two diverse fields, but also provides
a way to probe the NHTPs through nonequilibrium dynamics. To make
the connection more explicit, we will study the DQPTs in a couple
of prototypical 1D non-Hermitian lattice models in the following section.
Besides the rate function $g(t)$, we will also investigate the real-valued,
noncyclic geometric phase of the return amplitude $G(k,t)$~\cite{ZhouDQPT1},
which is defined as
\begin{equation}
\Phi_{{\rm G}}(k,t)=\Phi(k,t)-\Phi_{{\rm D}}(k,t),\label{eq:GP}
\end{equation}
where the total phase
\begin{equation}
\Phi(k,t)\equiv-i\ln\frac{G(k,t)}{|G(k,t)|},\label{eq:TP}
\end{equation}
and the dynamical phase
\begin{equation}
\Phi_{{\rm D}}(k,t)\equiv-\int_{0}^{t}dt'{\rm Re}\left\{ \frac{{\rm Tr}[\tilde{U}^{\dagger}(k,t')\rho_{0}U(k,t')H(k)]}{{\rm Tr}[\tilde{U}^{\dagger}(k,t')\rho_{0}U(k,t')]}\right\} \label{eq:DP}
\end{equation}
(see Appendix~\ref{sec:AppA} for more details about these phase factors).
The noncyclic geometric phase has been shown to contain important
information about DQPTs in both Hermitian~\cite{DTOP1,DTOP2,DTOP3,DTOP4,DTOP5} and non-Hermitian~\cite{ZhouDQPT1} systems. 
At a given time, the winding number of the geometric
phase in the first BZ can be further employed to
construct a dynamical topological order parameter (DTOP), which is defined
as
\begin{equation}
\nu(t)=\int\frac{dk}{2\pi}\partial_{k}\Phi_{{\rm G}}(k,t).\label{eq:DTOP}
\end{equation}
It takes a quantized jump whenever the evolution of the system
passes through a critical time of the DQPT. Note that the range of
integration over $k$ depends on the symmetry of $\Phi_{{\rm G}}(k,t)$
in $k$-space. For example, if $\Phi_{{\rm G}}(k,t)$ has the inversion
 symmetry with respect to $k=0$, i.e., $\Phi_{{\rm G}}(k,t)=\Phi_{{\rm G}}(-k,t)$,
we can perform the integral over a reduced BZ with $k\in[0,\pi]$ in the evaluation
of $\nu$ in Eq.~(\ref{eq:DTOP}). 

\section{Models and results}\label{sec:Model}

In this section, we demonstrate the connection between NHTPs and DQPTs
in three typical non-Hermitian 1D lattice models. In each subsection,
we introduce the model that will be investigated first and establish
its bulk topological phase diagram. After that, we consider the DQPTs
in the model following the quench from a trivial phase to different
non-Hermitian phases (either trivial or topological), and unveil the
relationship between the critical times and momenta of the DQPTs and
the topological invariants of the post-quench non-Hermitian system.
In the lossy Kitaev chain and its next-nearest-neighbor extension, we observe
a one-to-one correspondence between the NHTPs and DQPTs. In the nonreciprocal
SSH model, we find that while a topologically nontrivial post-quench
system always imply DQPTs following the quench, the reverse may not
be true in general, and possible reasons behind such an anomaly will
be discussed.

\subsection{The lossy Kitaev chain}\label{subsec:LKC}

We first consider a non-Hermitian variant of the Kitaev chain, which
describes a 1D topological superconductor with onsite particle loss.
In momentum representation, the Hamiltonian of the model takes the
form $H=\frac{1}{2}\sum_{k\in{\rm BZ}}\Psi_{k}^{\dagger}H(k)\Psi_{k}$,
where $\Psi_{k}^{\dagger}=(c_{k}^{\dagger},c_{-k})$ is the Nambu
spinor operator and $c_{k}^{\dagger}$ is the creation operator of
an electron with quasimomentum $k$. The Bloch Hamiltonian $H(k)$
in Nambu basis is given by 
\begin{equation}
H(k)=h_{y}(k)\sigma_{y}+[h_{z}(k)-{\rm i}v]\sigma_{z},\label{eq:HLKC}
\end{equation}
where
\begin{equation}
h_{y}(k)=\Delta\sin k,\qquad h_{z}(k)=u+J\cos k.\label{eq:HyzLKC}
\end{equation}
Here the real parameters $J$, $\Delta$ and $u$ denote the nearest-neighbor
hopping amplitude, superconducting pairing amplitude and chemical
potential. $v\in\mathbb{R}$ characterizes the strength of onsite
particle loss. Following the discussions of Subsec. \ref{subsec:NHTPs},
we see that $H(k)$ possesses the sublattice symmetry ${\cal S}=\sigma_{x}$,
i.e., ${\cal S}H(k){\cal S}=-H(k)$. Furthermore, it also has the
generalized particle-hole symmetry ${\cal C}=\sigma_{x}$ and time-reversal
symmetry ${\cal T}=\sigma_{0}$, in the sense that ${\cal C}H^{\top}(k){\cal C}^{-1}=-H(-k)$
and ${\cal T}H^{\top}(k){\cal T}^{-1}=H(-k)$, where $\top$ performs
matrix transposition. $H(k)$ thus belongs to an extension of the
symmetry class BDI in the periodic table of non-Hermitian topological
phases~\cite{NHTP3}. In the meantime, $H(k)$ possesses the inversion symmetry
${\cal P}=\sigma_{z}$ as ${\cal P}H(k){\cal P}^{-1}=H(-k)$, which
guarantees the correspondence between its bulk topological invariant
$w$~(as defined in Eq.~(\ref{eq:w})) and the number of Majorana
edge modes under the open boundary condition~\cite{NHTP3}.

\begin{figure}
	\begin{centering}
		\includegraphics[scale=0.49]{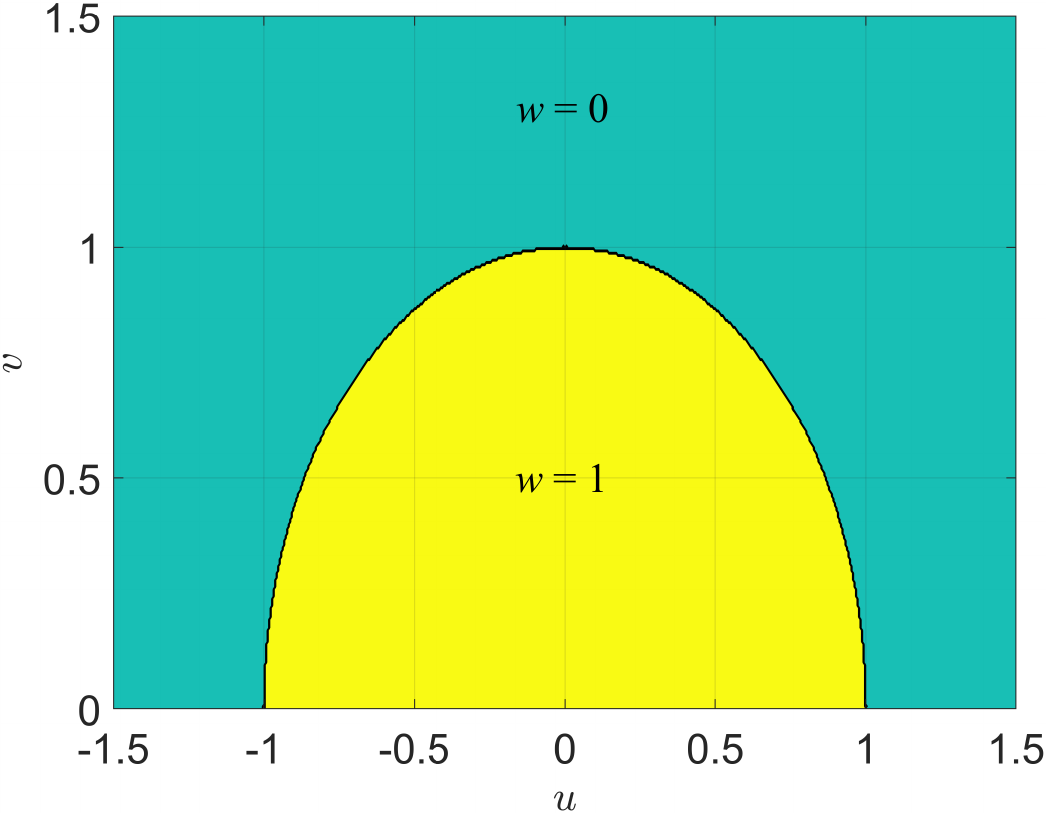}
		\par\end{centering}
	\caption{Topological phase diagram of the LKC versus the real and imaginary
		parts of chemical potential $u$ and $v$. Other system parameters
		are chosen as $J=\Delta=1$. Each region with a uniform color denotes
		a NHTP, with the value of winding number $w$ denoted explicitly therein.\label{fig:LKCPDs}}
\end{figure}

According to Subsec.~\ref{subsec:NHTPs}, the complex energy spectrum
of LKC takes the form
\begin{equation}
E_{\pm}(k)=\pm\sqrt{h_{y}^{2}(k)+[h_{z}(k)-iv]^{2}}=\pm E(k),\label{eq:EkLKC}
\end{equation}
which will become gapless when
\begin{alignat}{1}
\Delta\sin k & = \pm v,\label{eq:Gaples1LKC}\\
u+J\cos k & = 0.\label{eq:Gaples2LKC}
\end{alignat}
Combining these equations, we find the gapless quasimomenta 
\begin{equation}
\pm k_{0}=\pm\arccos(-u/J)\label{eq:k0LKC}
\end{equation}
for $|u|<|J|$, and the boundary between different NHTPs as
\begin{equation}
\frac{u^{2}}{J^{2}}+\frac{v^{2}}{\Delta^{2}}=1.\label{eq:PBsLKC}
\end{equation}
Geometrically, the trajectory of vector ${\bf h}(k)\equiv[h_{y}(k),h_{z}(k)]$
forms an ellipse on the $h_{y}$-$h_{z}$ plane, which is centered
at $(0,u)$. When the gapless condition Eq.~(\ref{eq:PBsLKC}) is
satisfied, the spectrum $E_{\pm}(k)$ hold a pair of EPs at $(\pm v,0)$
on the $h_{y}$-$h_{z}$ plane, which are passed through by the vector
${\bf h}(k)$. Whether the two EPs are encircled or not by the trajectory
of ${\bf h}(k)$ when $k$ scans through the first BZ then distinguishes
two possible NHTPs. With Eqs.~(\ref{eq:EkLKC}) and (\ref{eq:PBsLKC}),
it is not hard to show that when $u^{2}/J^{2}+v^{2}/\Delta^{2}<1$
($>1$), the two EPs are encircled (not encircled) by the trajectory
of ${\bf h}(k)$. The topological invariant that distinguish these
two phases has the form of Eq.~(\ref{eq:w}), where the winding angle
$\phi(k)$ for the LKC is explicitly given by
\begin{equation}
\phi(k)=\arctan\left[\frac{h_{z}(k)-iv}{h_{y}(k)}\right].\label{eq:PhikLKC}
\end{equation}
In Fig.~\ref{fig:LKCPDs}, we show the topological phase diagram of
the LKC versus the real and imaginary parts of chemical potential
$u$ and $v$, with $J=\Delta=1$. The winding numbers Eq.~(\ref{eq:w}) of
the non-Hermitian topological and trivial phases are found to be $w=1$
and $w=0$ for $u^{2}/J^{2}+v^{2}/\Delta^{2}<1$ and $>1$, respectively.
A loss-induced topological phase transition, which is unique to non-Hermitian
systems, can be observed with the increase of $v$.

\begin{figure}
	\begin{centering}
		\includegraphics[scale=0.5]{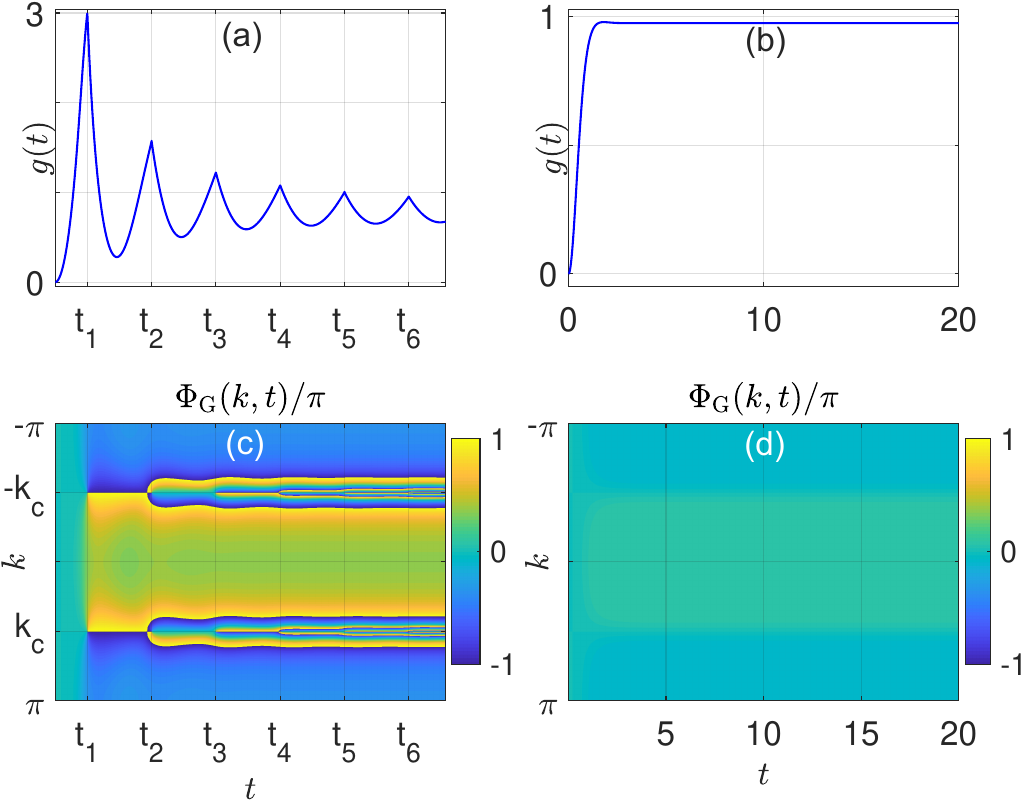}
		\par\end{centering}
	\caption{Rate function of return probability $g(t)$ \cite{Note1} and noncyclic
		geometric phase $\Phi_{{\rm G}}(k,t)$ of the LKC model. Panels (a) and
		(c) present the case with system parameters $J=\Delta=1$, $u=0$
		and $v=0.3$, in which the winding number $w=1$ (topological phase)
		and DQPTs are observed as the cusps in $g(t)$. The critical momenta
		$\pm k_{c}$ and critical times $t_{n}$ ($n=1,...,6$) are obtained
		from Eqs.~(\ref{eq:kcLKC}) and (\ref{eq:tnLKC}). An extra $2\pi$-jump
		of the geometric phase $\Phi_{{\rm G}}(k,t)$ is observed at $\pm k_{c}$
		every time when the evolution of the system passes through a critical
		time $t_{n}$. Panels (b) and (d) present the case with system parameters
		$J=\Delta=1$, $u=0$ and $v=1.3$, where there are no DQPTs and the
		winding number $w=0$ (trivial phase).\label{fig:LKCDQPTs}}
\end{figure}

To link the NHTPs of LKC with the DQPTs, we employ the protocol introduced
in Subsec.~\ref{subsec:DQPTs}, with the initial state $\rho_{0}=\sigma_{0}/2$
and the dynamics being governed by the Hamiltonian $H(k)$ in Eq.~(\ref{eq:HLKC}). 
According to Eq.~(\ref{eq:Gkt}), the return amplitude
at a later time $t>0$ is given by $G(k,t)=\cos[E(k)t]$, where $E(k)$
is the dispersion relation of LKC in Eq.~(\ref{eq:EkLKC}). From Eqs.~(\ref{eq:kc2})
and (\ref{eq:tn}), we find the critical momenta and times to be
\begin{alignat}{1}
\pm k_{c}= & \pm\arccos(-u/J)=\pm k_{0},\label{eq:kcLKC}\\
t_{n}(\pm k_{c})= & \left(n-\frac{1}{2}\right)\frac{\pi}{|\Delta|\sqrt{1-(u^{2}/J^{2}+v^{2}/\Delta^{2})}}.\label{eq:tnLKC}
\end{alignat}
It is clear that when $u^{2}/J^{2}+v^{2}/\Delta^{2}<1$, there are
real solutions of $t_{n}(k_{c})$ for all $n\in\mathbb{Z}$, and the
two critical momenta $\pm k_{c}$ are coincide with the gapless quasimomenta
$\pm k_{0}$, yielding the same critical period $T(k_{c})=\pi/\sqrt{\Delta^{2}(1-u^{2}/J^{2})-v^{2}}$.
On the other hand, there is no critical momenta and $t_{n}$ is always
imaginary when $u^{2}/J^{2}+v^{2}/\Delta^{2}>1$, yielding no DQPTs
at any real time $t$. When $u^{2}/J^{2}+v^{2}/\Delta^{2}=1$, which
corresponds to a gapless post-quench phase, we will have $t_{n}(k_{c})\rightarrow\infty$
for any solutions of critical momenta $\pm k_{c}$, and the resulting
DQPTs are not observable. For completeness, we numerically compute
the return rates and geometric phases with the help of Eqs.~(\ref{eq:gt})
and (\ref{eq:GP}) for the cases with and without DQPTs in Fig.~\ref{fig:LKCDQPTs}(a,c)
and \ref{fig:LKCDQPTs}(b,d), respectively. As expected, DQPTs are
only observed when the system is quenched to a nontrivial NHTP with
the winding number $w=1$.

Combining the analysis in this subsection, we obtain a \emph{one-to-one
	correspondence} between the NHTPs and DQPTs in the LKC, which is summarized
in Table \ref{tab:LKC}. This connection not only unifies the NHTPs
and DQPTs in the system, but also provides a way to dynamically distinguishing
the different NHTPs of LKC and detecting the gapless quasimomenta,
as exemplified by Figs.~\ref{fig:LKCDQPTs}(a,c).

\begin{table*}[t]
	\begin{centering}
		\begin{tabular}{cccc}
			\hline 
			\multirow{2}{*}{Condition} & Geometric & Winding & Critical time\tabularnewline
			& picture & number & and momenta\tabularnewline
			\hline 
			\hline 
			\multirow{2}{*}{$\frac{u^{2}}{J^{2}}+\frac{v^{2}}{\Delta^{2}}<1$} & Two EPs are & \multirow{2}{*}{$w=1$} & DQPTs at $t_{n}(k_{c})$ \tabularnewline
			& encircled by ${\bf h}(k)$ &  & $\forall n\in\mathbb{Z}$, $k_{c}=k_{0}$\tabularnewline
			\hline 
			\multirow{2}{*}{$\frac{u^{2}}{J^{2}}+\frac{v^{2}}{\Delta^{2}}=1$} & Two EPs are & \multirow{2}{*}{$/$} & \multirow{2}{*}{$/$}\tabularnewline
			& crossed by ${\bf h}(k)$ &  & \tabularnewline
			\hline 
			\multirow{2}{*}{$\frac{u^{2}}{J^{2}}+\frac{v^{2}}{\Delta^{2}}>1$} & No EPs are & \multirow{2}{*}{$w=0$} & No $k_{c}$ and $t_{n}$\tabularnewline
			& encircled by ${\bf h}(k)$ &  & No DQPTs\tabularnewline
			\hline 
		\end{tabular}
		\par\end{centering}
	\caption{The connection between NHTPs and DQPTs of the LKC model. The notation
		``$/$'' in the table means ``ill-defined''. $k_0$ refers to the gapless quasimomenta of the model.\label{tab:LKC}}
\end{table*}

\subsection{The lossy Kitaev chain with next-nearest-neighbor hoppings and pairings}\label{subsec:NNN-LKC}
\begin{figure}
	\begin{centering}
		\includegraphics[scale=0.5]{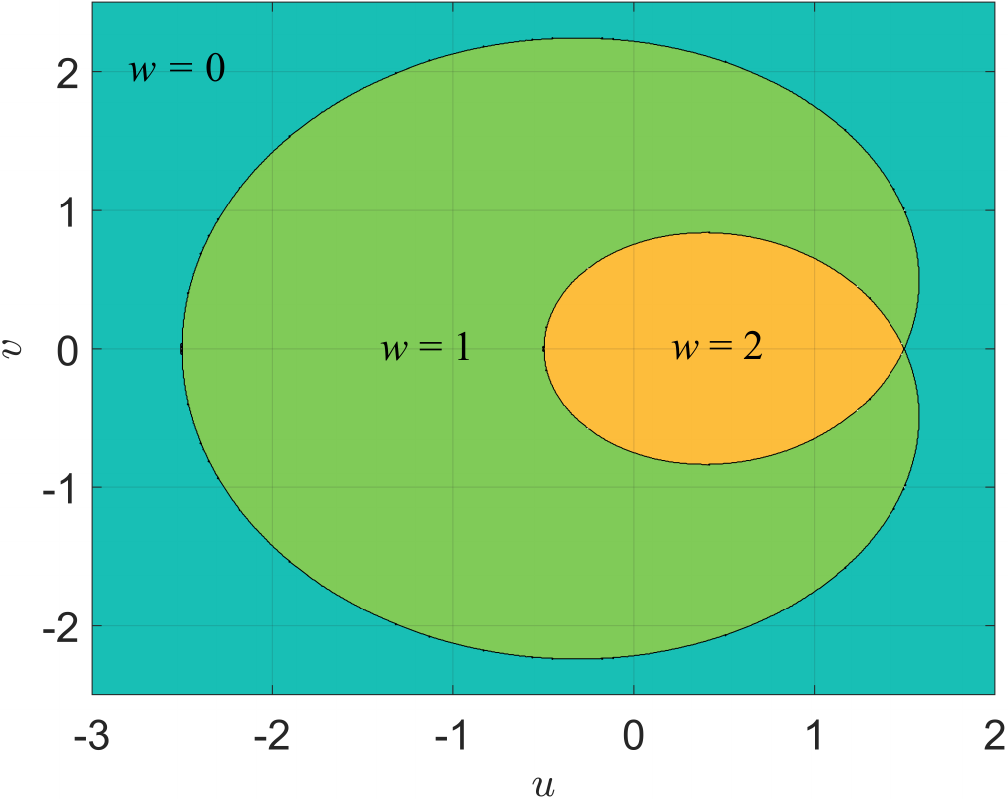}
		\par\end{centering}
	\caption{Topological phase diagram of the NNN LKC model versus the real and
		imaginary parts of chemical potential $u$ and $v$. Other system
		parameters are chosen as $J_{1}=\Delta_{1}=1$ and $J_{2}=\Delta_{2}=1.5$.
		Each region with a uniform color corresponds to a NHTP, with the
		value of topological winding number $w$ denoted explicitly therein.
		The black line separating different regions is the phase boundary
		obtained from Eq.~(\ref{eq:PBsLKC-NNN}).\label{fig:NNN-LKCPDs}}
\end{figure}

We next consider the LKC with NNN hoppings and pairings, which could
possesses NHTPs with larger topological invariants. In the momentum
space and Nambu spinor basis, the NNN LKC is described by the Hamiltonian
$H=\frac{1}{2}\sum_{k\in{\rm BZ}}\Psi_{k}^{\dagger}H(k)\Psi_{k}$,
where $H(k)$ takes the same form as Eq.~(\ref{eq:HLKC}), with
\begin{alignat}{1}
h_{y}(k) &= \Delta_{1}\sin k+\Delta_{2}\sin2k,\nonumber \\
h_{z}(k) &= u+J_{1}\cos k+J_{2}\cos2k.\label{eq:HyzLKC-NNN}
\end{alignat}
Here $u$ is the real part of chemical potential, $(J_{1},\Delta_{1})$ and 
$(J_{2},\Delta_{2})$ are the nearest-neighbor and next-nearest-neighbor hopping
and pairing amplitudes, respectively. It is not hard to verify that the $H(k)$
here belongs to the same symmetry class as the LKC, with the same
set of time-reversal, particle-hole, sublattice and inversion symmetries.
The dispersion relations $E_{\pm}(k)$ of $H(k)$ share the same form
with Eq.~(\ref{eq:EkLKC}), yielding the gapless conditions
\begin{alignat}{1}
\Delta_{1}\sin k+\Delta_{2}\sin2k &= \pm v,\label{eq:Gaples1LKC-NNN}\\
u+J_{1}\cos k+J_{2}\cos2k &= 0.\label{eq:Gaples2LKC-NNN}
\end{alignat}
By solving Eq.~(\ref{eq:Gaples2LKC-NNN}), we could obtain at most
four possible gapless quasimomenta $\pm k_{0}^{\pm}$ as 
\begin{equation}
\pm k_{0}^{\pm}=\pm\arccos\left[\frac{-J_{1}\pm\sqrt{J_{1}^{2}+8J_{2}(J_{2}-u)}}{4J_{2}}\right].\label{eq:k0LKC-NNN}
\end{equation}
According to Eq.~(\ref{eq:Gaples1LKC-NNN}), the boundary between
different NHTPs is then determined by
\begin{equation}
\sin k_{0}^{\pm}(\Delta_{1}+2\Delta_{2}\cos k_{0}^{\pm})=\pm v.\label{eq:PBsLKC-NNN}
\end{equation}
The explicit expression of the phase boundary in terms of system parameters
is tedious, and will be left for numerical calculations. Geometrically,
the trajectory of real vector ${\bf h}(k)=[h_{y}(k),h_{z}(k)]$ has
the shape of a centered trochoid, which could encircle the two EPs
of $E(k)$ at $(\pm v,0)$ twice, once or zero times when $k$ is
scanned over the first BZ. These three possibilities then distinguish
three different NHTPs, which are characterized by the topological
winding number $w$ in Eq.~(\ref{eq:w}). In Fig.~\ref{fig:NNN-LKCPDs},
we present the topological phase diagram of the NNN LKC model versus
the real and imaginary parts of chemical potential $u$ and $v$,
with other system parameters set as $J_{1}=\Delta_{1}=1$ and $J_{2}=\Delta_{2}=1.5$.
The three topological phases are discriminated by different colored
regions in the figure, with the phase boundary curve (black solid
line) determined by Eq.~(\ref{eq:PBsLKC-NNN}), and the value of $w$
denoted explicitly within each phase. The quantized changes of $w$
with the increase of the lossy strength $v$ again signify non-Hermiticity
induced topological phase transitions in the system.

To build the connection between the NHTPs and the DQPTs of NNN LKC,
we again employ the protocol introduced in Subsec.~\ref{subsec:DQPTs}, with
the initial state $\rho_{0}=\sigma_{0}/2$ and the dynamics being
governed by the Hamiltonian $H(k)$ of the NNN LKC. The return amplitude
at a later time $t>0$ is then given by $G(k,t)=\cos[E(k)t]$, with
$E(k)$ being the dispersion of NNN LKC. The critical momenta and
time are further obtained from Eqs.~(\ref{eq:kc2}) and (\ref{eq:tn})
as
\begin{alignat}{1}
\pm k_{c}^{\pm}= & \pm\arccos\left[\frac{-J_{1}\pm\sqrt{J_{1}^{2}+8J_{2}(J_{2}-u)}}{4J_{2}}\right]=\pm k_{0}^{\pm},\label{eq:kcLKC-NNN}\\
t_{n}(k_{c}^{\pm})= & \left(n-\frac{1}{2}\right)\frac{\pi}{\sqrt{(\Delta_{1}\sin k_{c}^{\pm}+\Delta_{2}\sin2k_{c}^{\pm})^{2}-v^{2}}},\label{eq:tnLKC-NNN}
\end{alignat}
where $n\in\mathbb{Z}$. In parallel with the discussions of Subsec.~\ref{subsec:LKC}, 
we could summarize the relationship between NHTPs
and DQPTs in the NNN LKC model by Table \ref{tab:NNN-LKC}. Again,
we obtain a one-to-one correspondence between these two nonequilibrium
phenomena, which also provides us with a way to detect NHTPs with
large winding numbers and to locate the phase boundaries between them.

\begin{table*}[t]
	\begin{centering}
		\begin{tabular}{cccc}
			\hline 
			\multirow{2}{*}{Condition} & Geometric & Winding & Critical times\tabularnewline
			& picture & number & and momenta\tabularnewline
			\hline 
			\hline 
			\multirow{2}{*}{$h_{y}^{2}(k_{c}^{\pm})>v^{2}$} & Two EPs are encircled & \multirow{2}{*}{$w=2$} & DQPTs at $t_{n}(k_{c}^{\pm})$ \tabularnewline
			& twice by ${\bf h}(k)$ &  & $\forall n\in\mathbb{Z}$, $k_{c}^{\pm}=k_{0}^{\pm}$\tabularnewline
			\hline 
			$h_{y}^{2}(k_{c}^{+/-})>v^{2}$ & Two EPs are encircled & \multirow{2}{*}{$w=1$} & DQPTs at $t_{n}(k_{c}^{+/-})$ \tabularnewline
			\& $h_{y}^{2}(k_{c}^{-/+})<v^{2}$ & once by ${\bf h}(k)$ &  & $\forall n\in\mathbb{Z}$, $k_{c}^{\pm}=k_{0}^{\pm}$\tabularnewline
			\hline 
			\multirow{2}{*}{$h_{y}^{2}(k_{c}^{\pm})<v^{2}$} & No EPs are encircled & \multirow{2}{*}{$w=0$} & No $k_{c}$ and $t_{n}$\tabularnewline
			& by ${\bf h}(k)$ &  & No DQPTs\tabularnewline
			\hline 
		\end{tabular}
		\par\end{centering}
	\caption{The connection between NHTPs and DQPTs of the NNN LKC model.\label{tab:NNN-LKC}}
\end{table*}

\begin{figure}
	\begin{centering}
		\includegraphics[scale=0.5]{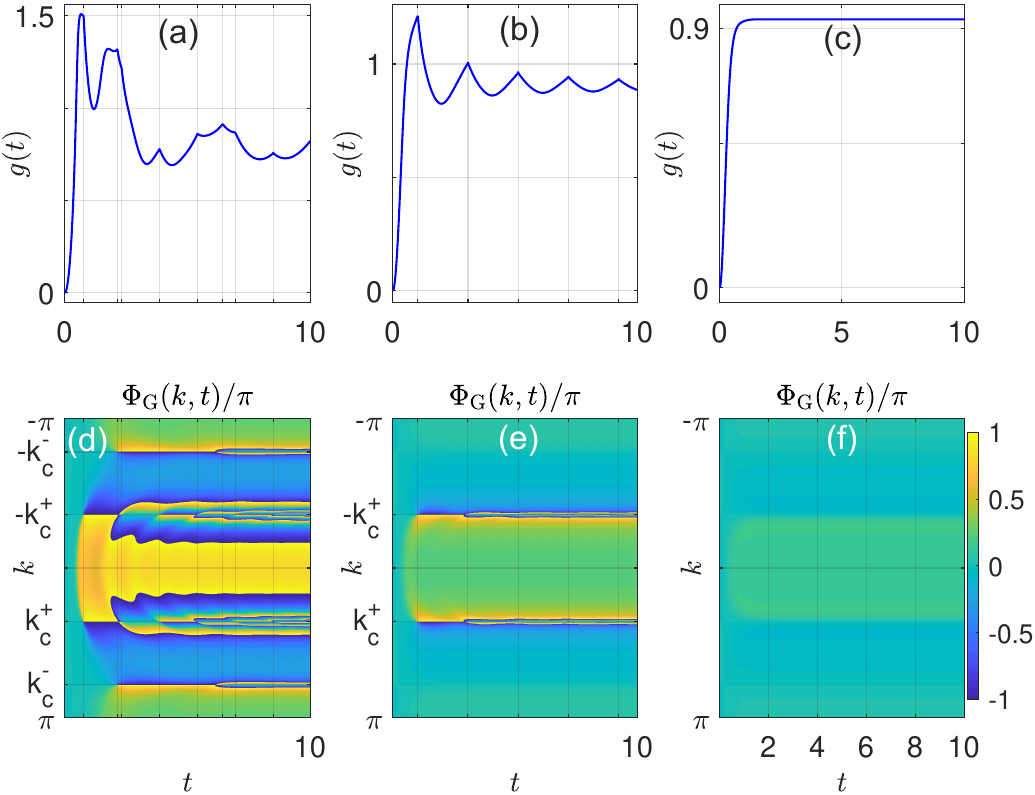}
		\par\end{centering}
	\caption{The rate function $g(t)$ in panels (a,b,c), and geometric
		phase $\Phi_{{\rm G}}(t)$ in panels (d,e,f) of the NNN LKC
		model~\cite{Note1}. The system parameters for the post-quench Hamiltonian are $J_{1}=\Delta_{1}=1$,
		$J_{2}=\Delta_{2}=1.5$, $u=0.5$, $v=0.4$, $1.4$ and $2.4$ in
		panels (a,d), (b,e) and (c,f), respectively. The winding numbers are
		$w=2,1$ and $0$ for the cases in panels (a,d), (b,e) and (c,f).
		DQPTs are observed as the cusps in $g(t)$ in panels (a) and (b).
		In panels (a,d), the ticks along the horizontal axis denote the critical
		times $t_{1}(k_{c}^{+})$, $t_{1}(k_{c}^{-})$, $t_{2}(k_{c}^{+})$,
		$t_{3}(k_{c}^{+})$, $t_{4}(k_{c}^{+})$, $t_{2}(k_{c}^{-})$, $t_{5}(k_{c}^{+})$,
		$t_{6}(k_{c}^{+})$ from left to right, whose explicit values are
		obtained from Eqs.~(\ref{eq:kcLKC-NNN}) and (\ref{eq:tnLKC-NNN}).
		In panels (b,e), the ticks along the horizontal axis are the critical
		times $t_{n}(k_{c}^{+})$ for $n=1,...,5$ from left to right. An
		extra amount of $2\pi$-jump in the geometric phase $\Phi_{{\rm G}}(k,t)$
		is observed at the corresponding critical momentum when a DQPT
		happens.\label{fig:NNN-LKCDQPTs}}
\end{figure}

For completeness, we present three numerical examples for the DQPTs
in the NNN LKC in Fig.~\ref{fig:NNN-LKCDQPTs}. The system parameters
are chosen as $J_{1}=\Delta_{1}=1$, $J_{2}=\Delta_{2}=1.5$, $u=0.5$
and $v=0.4,1.4,2.4$ for figure panels \ref{fig:NNN-LKCDQPTs}(a,d),
\ref{fig:NNN-LKCDQPTs}(b,e) and \ref{fig:NNN-LKCDQPTs}(c,f). In
Figs.~\ref{fig:NNN-LKCDQPTs}(a,d) the post-quench system is in a
NHTP with $w=2$, and DQPTs are observed at two different sets of
critical periods $T(k_{c}^{\pm})=\pi/\sqrt{h_{y}^{2}(k_{c}^{\pm})-v^{2}}$
of $g(t)$ in Fig.~\ref{fig:NNN-LKCDQPTs}(a), with the two pairs
of critical momenta $\pm k_{c}^{\pm}$ given by Eq.~(\ref{eq:kcLKC-NNN})
and imaged by the $2\pi$-jumps of geometric phase $\Phi_{{\rm G}}(k,t)$
in Fig.~\ref{fig:NNN-LKCDQPTs}(d). In Figs.~\ref{fig:NNN-LKCDQPTs}(b,e),
the post-quench system is in a NHTP with $w=1$, and DQPTs are repeated
at only one critical period $T(k_{c}^{+})=\pi/\sqrt{h_{y}^{2}(k_{c}^{+})-v^{2}}$
of $g(t)$ in Fig.~\ref{fig:NNN-LKCDQPTs}(b), with $2\pi$-jumps
of geometric phase $\Phi_{{\rm G}}(k,t)$ observed at the critical
momenta $\pm k_{c}^{+}$ in Fig.~\ref{fig:NNN-LKCDQPTs}(e). In Figs.~\ref{fig:NNN-LKCDQPTs}(c,f), 
the post-quench system is in a trivial
phase with $w=0$, and no signatures of DQPTs are observed in the
rate function $g(t)$ and geometric phase $\Phi_{{\rm G}}(k,t)$.
Putting together, our numerical results confirm the connection between
the NHTPs and DQPTs of the NNN LKC model, as summarized in Table \ref{tab:NNN-LKC}.
Furthermore, the results presented here should be directly extendable
to non-Hermitian models in the same symmetry class as the LKC, but
with even longer-range hopping and pairing amplitudes.

\subsection{The nonreciprocal SSH model}\label{subsec:NRSSH}
In the last part of this section, we consider a nonreciprocal variant
of the SSH model, which possesses a different set of symmetries compared
with the LKC. In momentum representation, the Hamiltonian of NRSSH
model takes the form $H=\sum_{k\in{\rm BZ}}\Psi_{k}^{\dagger}H(k)\Psi_{k}$,
where $\Psi_{k}^{\dagger}=(a_{k}^{\dagger},b_{k}^{\dagger})$ is the
creation operator on the two sublattices $a$ and $b$ of the SSH
model, and $k\in[-\pi,\pi)$ is the quasimomentum. The Bloch Hamiltonian
$H(k)$ is explicitly given by
\begin{equation}
H(k)=h_{x}(k)\sigma_{x}+[h_{y}(k)-i\gamma]\sigma_{y},\label{eq:HNRSSH}
\end{equation}
with
\begin{equation}
h_{x}(k)=J_{1}+J_{2}\cos k,\qquad h_{y}(k)=J_{2}\sin k.\label{eq:HxyNRSSH}
\end{equation}
Here $J_{1}\pm\gamma$ and $J_{2}$ are the intracell and intercell
hopping amplitudes. A finite $\gamma$ makes the intracell hopping
asymmetric, leading to a non-Hermitian $H(k)$. From now on, we assume
$J_{2},\gamma>0$ without loss of generality. It is clear that the
system possesses the sublattice symmetry ${\cal S}=\sigma_{z}$, in
the sense that ${\cal S}H(k){\cal S}=-H(k)$. This allows us to characterize
the bulk topological phases of $H(k)$ by the winding number $w$.
Moreover, $H(k)$ has the time reversal symmetry ${\cal T}=\sigma_{0}$
and particle-hole symmetry ${\cal C}=\sigma_{z}$, i.e., ${\cal T}H^{*}(k){\cal T}^{-1}=H(-k)$
and ${\cal C}H^{*}(k){\cal C}^{-1}=-H(-k)$. Therefore, the NRSSH
model belongs to the same BDI symmetry class as the Hermitian SSH
model. Nevertheless, $H(k)$ in Eq.~(\ref{eq:HNRSSH}) does not have
the inversion symmetry of the Hermitian SSH model, but instead possesses
the PT-symmetry, i.e., ${\cal PT}H^{*}(k)({\cal PT})^{-1}=H(k)$.
This allows the bulk spectrum of $H(k)$ to be very different under
periodic and open boundary conditions, leading to the breakdown of
conventional bulk-boundary correspondence~\cite{NHSkin1}.

\begin{figure}
	\begin{centering}
		\includegraphics[scale=0.49]{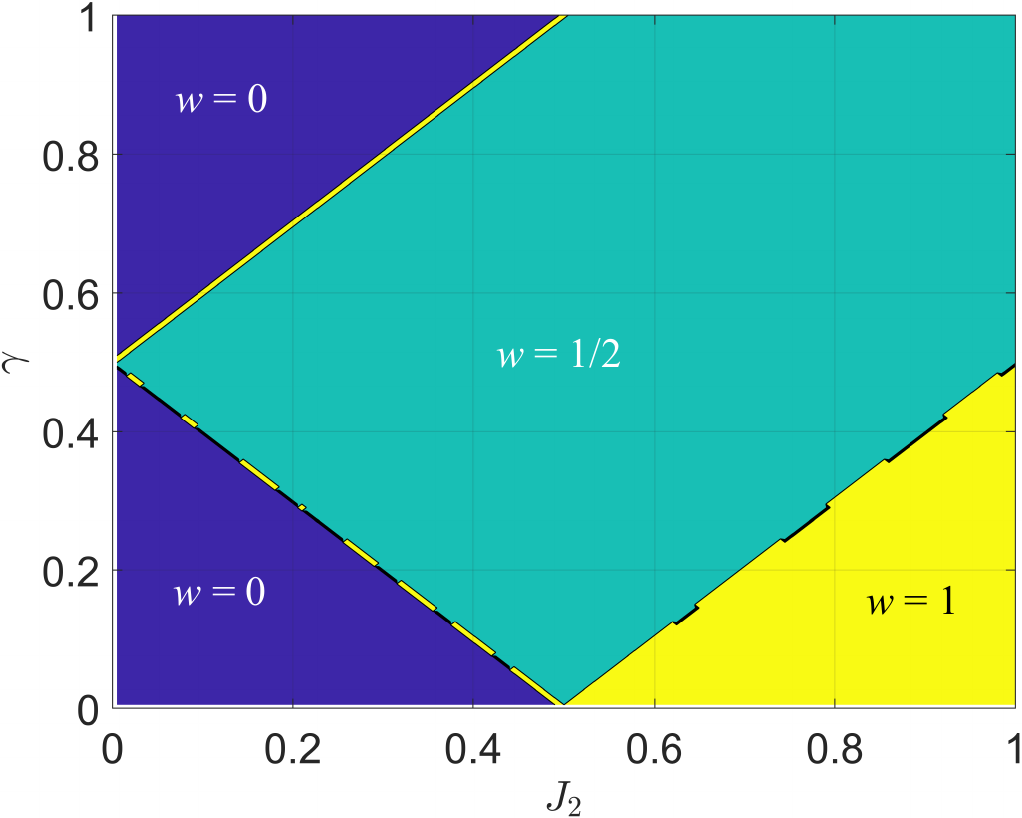}
		\par\end{centering}
	\caption{Topological phase diagram of the NRSSH model versus the intercell
		hopping amplitude $J_{2}$ and the asymmetric intracell modulation $\gamma$.
		The symmetric intracell hopping amplitude is set as $J_{1}=0.5$.
		Each regime with a uniform color corresponds to a NHTP, within which
		the topological winding number $w$ takes a constant value as shown
		in the figure. The lines separating different NHTPs are phase boundaries
		obtained from Eq.~(\ref{eq:PBsNRSSH}).\label{fig:NRSSHPDs}}
\end{figure}

The bulk spectrum of $H(k)$ takes the form
\begin{equation}
E_{\pm}(k)=\pm\sqrt{h_{x}^{2}(k)+[h_{y}(k)-i\gamma]^{2}}=\pm E(k).\label{eq:EkNRSSH}
\end{equation}
With Eq.~(\ref{eq:HxyNRSSH}), we see that the dispersion is gapless
at zero energy when the following two conditions are met
\begin{alignat}{1}
J_{1}\pm J_{2} &= \pm\gamma,\label{eq:PBsNRSSH}\\
\sin k &= 0,\label{eq:k0NRSSH}
\end{alignat}
which directly yield the phase boundary curves and the gapless quasimomenta
$k_{0}=0,\pi$. Geometrically, the trajectory of vector ${\bf h}(k)=[h_{x}(k),h_{y}(k)]$
forms a circle with radius $J_{2}$ and centered at $(J_{1},0)$ on
the $h_{x}$-$h_{y}$ plane, while the EPs of the spectrum are located
at $(\pm\gamma,0)$. When $J_{1}-J_{2}<-\gamma$ and $J_{1}+J_{2}>\gamma$,
both two EPs are encircled by ${\bf h}(k)$ when $k$ scans over the
first BZ. When $J_{1}-J_{2}<-\gamma$ ($J_{1}+J_{2}>\gamma$) and
$|J_{1}+J_{2}|<\gamma$ ($|J_{1}-J_{2}|<\gamma$), the EP at $(-\gamma,0)$
($(\gamma,0)$) is encircled by ${\bf h}(k)$. Otherwise no EPs
are encircled by ${\bf h}(k)$. These three different situations distinguish
three different types of bulk non-Hermitian topological phases, with
each of them being characterized by the topological invariant $w$
in Eq.~(\ref{eq:w}), where the winding angle
\begin{equation}
\phi(k)=\arctan\left[\frac{h_{y}(k)-i\gamma}{h_{x}(k)}\right].
\end{equation}
In Fig.~\ref{fig:NRSSHPDs}, we present the topological phase diagram
of the NRSSH model versus the intercell hopping amplitude $J_{2}$
and asymmetric hopping parameter $\gamma$, with the intracell hopping
amplitude $J_{1}=0.5$. Each colored region in the phase diagram corresponds
to a NHTP, with the value of topological winding number $w$ denoted
therein. The phase boundaries separating different regions are determined
by Eq.~(\ref{eq:PBsNRSSH}). Despite non-Hermiticity-induced topological
phase transitions, the NRSSH model also features a unique NHTP with winding
number $w=1/2$, which corresponds to the case in which only a single
EP is encircled by ${\bf h}(k)$. 

The connection between DQPTs and NHTPs in the NRSSH model can be built
as follows. Choosing the initial state to be $\rho_{0}=\sigma_{0}/2$
as in Subsec.~\ref{subsec:DQPTs}, the dynamics of the system at $t>0$
is governed by the Hamiltonian $H(k)$ of the NRSSH model. The return
amplitude at time $t$ is given by $G(k,t)=\cos[E(k)t]$, with $E(k)$
being the dispersion of $H(k)$. The critical momenta and time are
then obtained from Eqs.~(\ref{eq:kc2}) and (\ref{eq:tn}), i.e.,
\begin{alignat}{1}
k_{c} &= (0,\pi)=k_{0},\label{eq:kcNRSSH}\\
t_{n}(0) &= \left(n-\frac{1}{2}\right)\frac{\pi}{\sqrt{(J_{1}+J_{2}-\gamma)(J_{1}+J_{2}+\gamma)}}\equiv t_{n}^{0},\label{eq:t0NRSSH}\\
t_{n}(\pi) &= \left(n-\frac{1}{2}\right)\frac{\pi}{\sqrt{(J_{1}-J_{2}-\gamma)(J_{1}-J_{2}+\gamma)}}\equiv t_{n}^{\pi},\label{eq:tpNRSSH}
\end{alignat}
where $n\in\mathbb{Z}$. Combining these equations with the gapless
conditions in Eqs.~(\ref{eq:PBsNRSSH}) and (\ref{eq:k0NRSSH}), we could
immediately identify the relationship between NHTPs and DQPTs in the NRSSH
model, as listed in Table~\ref{tab:NRSSH}.

\begin{table*}[t]
	\begin{centering}
		\begin{tabular}{cccc}
			\hline 
			\multirow{2}{*}{Condition} & Geometric & Winding & Critical times\tabularnewline
			& picture & number & and momenta\tabularnewline
			\hline 
			\hline 
			$J_{1}-J_{2}<-\gamma$ & Two EPs are & \multirow{2}{*}{$w=1$} & DQPTs at $t_{n}^{0,\pi}$ \tabularnewline
			\& $J_{1}+J_{2}>\gamma$ & encircled by ${\bf h}(k)$ &  & $\forall n\in\mathbb{Z}$, $k_{c}=0,\pi$\tabularnewline
			\hline 
			$J_{1}-J_{2}<-\gamma$ &  & \multirow{4}{*}{$w=1/2$} & DQPTs at $t_{n}^{\pi}$ \tabularnewline
			\& $|J_{1}+J_{2}|<\gamma$ & One EP is &  & $\forall n\in\mathbb{Z}$, $k_{c}=\pi$\tabularnewline
			\cline{1-1} \cline{4-4} 
			$J_{1}+J_{2}>\gamma$ & encircled by ${\bf h}(k)$ &  & DQPTs at $t_{n}^{0}$ \tabularnewline
			\& $|J_{1}-J_{2}|<\gamma$ &  &  & $\forall n\in\mathbb{Z}$, $k_{c}=0$\tabularnewline
			\hline 
			\multirow{2}{*}{$|J_{1}\pm J_{2}|<\gamma$} &  & \multirow{4}{*}{$w=0$} & No $k_{c}$ and $t_{n}$\tabularnewline
			& No EPs are &  & No DQPTs\tabularnewline
			\cline{1-1} \cline{4-4} 
			\multirow{2}{*}{$|J_{1}\pm J_{2}|>\gamma$} & encircled by ${\bf h}(k)$ &  & DQPTs at $t_{n}^{0,\pi}$\tabularnewline
			&  &  & $\forall n\in\mathbb{Z}$, $k_{c}=0,\pi$\tabularnewline
			\hline 
		\end{tabular}
		\par\end{centering}
	\caption{The connection between NHTPs and DQPTs of the NRSSH model.\label{tab:NRSSH}}
\end{table*}

From the table, we observe that since there is only a single critical
momentum for the NHTPs with $w=1/2$, there is also a unique set of
critical times ($t_{n}^{0}$ or $t_{n}^{\pi}$ for $n\in\mathbb{Z}$)
for the DQPTs in this case. This is in contrast with the NHTPs having
$w=1$, for which both $k_{c}=k_{0}=0$ and $\pi$ are the critical momenta,
and DQPTs at two different critical time periods $T(k_{c}=0,\pi)$
are expected in the post-quench dynamics. In the meantime, we also
observe an anomalous case as shown in the last row of
Table \ref{tab:NRSSH}. In this case, DQPTs are found when the post-quench
system is in a trivial phase with $w=0$. Therefore, even though a
nontrivial topological phase of the NRSSH model always lead to a
unique set of DQPTs following the quench to that phase, the reverse
is not true in general. Such a breakdown of the one-to-one correspondence
between the DQPTs and NHTPs in the NRSSH model might be due to the
absence of inversion symmetry, as compared with the situations in
the LKC and its NNN extension. Nevertheless, the most intriguing phase
of the NRSSH model, i.e., the one with $w=1/2$ can still be distinguished
from the other phases through the DQPTs. Therefore, the connection
between NHTPs and DQPTs we discovered can still be used as a powerful
tool to probe the details of the NHTPs in the NRSSH model.

\begin{figure}
	\begin{centering}
		\includegraphics[scale=0.5]{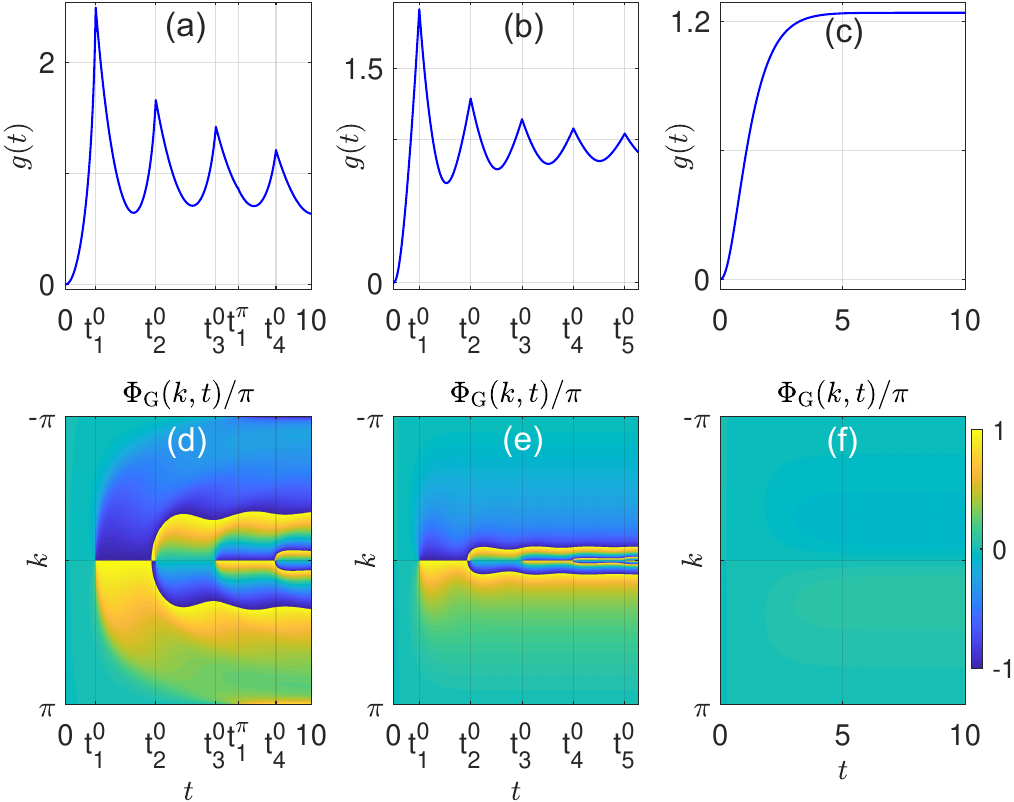}
		\par\end{centering}
	\caption{The rate function $g(t)$~\cite{Note1} in panels (a,b,c), and geometric
		phase $\Phi_{{\rm G}}(t)$ in panels (d,e,f) of the NRSSH model.
		The system parameters for the post-quench Hamiltonian are $J_{1}=0.5$,
		$(J_{2},\gamma)=(0.8,0.2)$, $(0.4,0.5)$ and $(0.2,0.8)$ in panels
		(a,d), (b,e) and (c,f), respectively. The winding numbers are $w=1,1/2$
		and $0$ for the cases in panels (a,d), (b,e) and (c,f). DQPTs are
		observed as the cusps in $g(t)$ in panels (a) and (b). In panels
		(a,b,d,e), the ticks along the horizontal axis denote the critical
		times from left to right, whose explicit values are obtained from
		Eqs.~(\ref{eq:kcNRSSH})-(\ref{eq:tpNRSSH}). An extra amount of $2\pi$-jump
		in the geometric phase $\Phi_{{\rm G}}(k,t)$ is observed around the
		corresponding critical momentum in panels (d,e) when a DQPT happens.\label{fig:NRSSHDQPTs}}
\end{figure}

For completeness, we present the DQPTs in three typical post-quench
phases of the NRSSH model in Fig.~\ref{fig:NRSSHDQPTs}. In Figs.~\ref{fig:NRSSHDQPTs}(a,d), 
the post-quench phase has winding number
$w=1$, and DQPTs are observed as cusps in the rate function $g(t)$
at two sets of critical times $t_{n}^{0,\pi}$. At each critical time,
a $2\pi$-jump in the geometric phase $\Phi_{{\rm G}}(k,t)$ is observed
at both the critical momenta $k_{c}=0,\pi$. In Figs.~\ref{fig:NRSSHDQPTs}(b,e),
the post-quench phase has winding number $w=1/2$, and DQPTs are found
at a unique set of critical time $t_{n}^{0}$, where a $2\pi$-jump
in the geometric phase $\Phi_{{\rm G}}(k,t)$ is observed around $k_{c}=0$.
In Figs.~\ref{fig:NRSSHDQPTs}(c,f), the post-quench phase
is trivial and no DQPTs are found in the post-quench dynamics. Putting
together, we found that the NHTPs and DQPTs in the NRSSH model are
also two closely related phenomena, and the later can be employed
to dynamically probe the properties of the former.

\section{Experimental proposal}\label{sec:Exp}

With all the theoretical and numerical results presented above, we
now sketch an experimental proposal in which our predicted connection
between NHTPs and DQPTs may be verified. Recently, a setup containing
an NV center in diamond has been employed to realize the PT-symmetry
breaking transition of a non-Hermitian two-level Hamiltonian~\cite{NVExp0}.
The general idea is to dilate a PT-symmetric Hamiltonian into a Hermitian
one, and execute the dynamics with the dilated Hamiltonian. Since
all the three models discussed in the previous section possess two
bulk bands together with the PT-symmetry, the setup proposed in Ref.~\cite{NVExp0} 
tends out to be an ideal platform in which the topological
invariants and DQPTs of our systems can be detected.

The Hamiltonian we are interested in, as shown in Eq.~(\ref{eq:Hk})
can be generally expressed as $H(k)={\bf d}(k)\cdot\boldsymbol{\sigma}$,
where ${\bf d}(k)=[h_{a}(k)-ig_{a}(k),h_{b}(k)-ig_{b}(k)]$ and $\boldsymbol{\sigma}=(\sigma_{a},\sigma_{b})$.
The strategy of Ref.~\cite{NVExp0} is to dilate $H(k)$ into a Hermitian
counterpart with the help of an ancilla qubit. The dilated Hamiltonian
$H'(k,t)$ yields the Schr\"odinger equation
\begin{equation}
i\frac{d}{dt}|\Omega(k,t)\rangle=H'(k,t)|\Omega(k,t)\rangle,\label{eq:Seq-Hd}
\end{equation}
where $|\Omega(k,t)\rangle$ denotes the state of the composite system.
With an appropriate post-selection scheme, the measurement results
can be restricted to a unique outcome for the ancilla qubit~\cite{NVExp0}.
Within the scheme, the composite state $|\Omega(k,t)\rangle$ takes
the form
\begin{equation}
|\Omega(k,t)\rangle=|\Psi(k,t)\rangle|-\rangle+\omega(t)|\Psi(k,t)\rangle|+\rangle,\label{eq:dState}
\end{equation}
where $\omega(t)$ is an appropriate linear operator, and the ancilla
qubit basis $|\pm\rangle$ are chosen to be the eigenstates of Pauli matrix $\sigma_{y}$
\begin{equation}
|-\rangle=\frac{|0\rangle-i|1\rangle}{\sqrt{2}},\qquad|+\rangle=-i\frac{|0\rangle+i|1\rangle}{\sqrt{2}}.\label{eq:SyBasis}
\end{equation}
In the experiment, a $-\pi/2$ pulse is applied following the evolution,
and only the measurement results inside the state manifold $|\Psi(k,t)\rangle|-\rangle$
is post-selected.

The explicit form of dilated Hamiltonian $H'(k,t)$ is not unique.
A convenient choice realized by the experiment in Ref.~\cite{NVExp0}
is
\begin{equation}
H'(k,t)=\Lambda(k,t)\otimes\sigma_{0}+\Gamma(k,t)\otimes\sigma_{z},\label{eq:Hdkt}
\end{equation}
where 
\begin{equation}
\Lambda(k,t)=\left\{ H(k)+\left[i\frac{d}{dt}\omega(t)+\omega(t)H(k)\right]\omega(t)\right\} M^{-1}(t),\label{eq:Lamkt}
\end{equation}
\begin{equation}
\Gamma(k,t)=i\left[H(k)\omega(t)-\omega(t)H(k)-i\frac{d}{dt}\omega(t)\right]M^{-1}(t),\label{eq:Gamkt}
\end{equation}
with the time-dependent operator $M(t)\equiv\omega^{\dagger}(t)\omega(t)+\sigma_{0}$.
Expanding $\Lambda(k,t)$ and $\Gamma(k,t)$ by the Pauli matrices
$\sigma_{x,y,z}$ and $\sigma_{0}$, we can further express $H'(k,t)$
as
\begin{equation}
H'(k,t)=\sum_{i=0}^{3}\sigma_{i}\otimes\left[A_{i}(k,t)\sigma_{0}+B_{i}(k,t)\sigma_{z}\right],\label{eq:HdktAB}
\end{equation}
where the real coefficients $A_{i}(k,t)$ and $B_{i}(k,t)$ for $i=0,1,2,3$
can be obtained numerically~\cite{NVExp0}.

Experimentally, the dilated Hamiltonian $H'(k,t)$ contain four levels
at each $k$, which can be encoded in the ground state manifold of
electron and nuclear spins in an NV center. The dynamics of the system,
in which $H(k)$ takes the form of Eq.~(\ref{eq:HLKC}) or (\ref{eq:HNRSSH})
can be monitored in the population of post-selected state, which further
provides us with the information about DQPTs in the corresponding
lattice model. The topological winding numbers of the model can also
be obtained by measuring the dynamic winding number of time-averaged
spin textures~\cite{TAST}, as suggested in Ref.~\cite{DWN}. In a very recent
experiment, the non-Hermitian topological phases of a nonreciprocal SSH model
have been detected in an NV center setup following the universal dilation scheme~\cite{NVExp1}, which confirms the applicability of the experimental proposal.

\section{Summary}\label{sec:Sum}

In this manuscript, we establish a relationship between NHTPs and
DQPTs in 1D systems. DQPTs are found when the system is quenched from
a trivial to a non-Hermitian topological phase. The numbers
of critical momenta and the periods of critical time are further
related to the topological invariants of the post-quench non-Hermitian
phases. Our results are demonstrated explicitly in three characteristic
non-Hermitian lattice models, which possess non-Hermiticity induced
topological phase transitions. Finally, we introduce a proposal to
observe the connection between NHTPs and DQPTs by manipulating an
NV center in diamond. This work therefore bridges the gap between
two classes of fascinating nonequilibrium phenomena, the NHTPs and
DQPTs, and brings new insights about the dynamical characterization
of non-Hermitian states of matter.

In this work, our theory is applied to one-dimensional two-band models with chiral symmetry. Our initial attempts also suggest that the theoretical framework presented here is generalizable to chiral-symmetric multiple-band models~\cite{ZhouMCD2,NHSSH4}. However, due to the complexity of multiple-band systems in the study of NHTPs and DQPTs, we expect that our theory would subject to appropriate modifications when it is applied to these systems. This interesting topic will be left for future explorations. In the meantime, it would be interesting to extend our findings to non-Hermitian systems under open boundary conditions, where the non-Hermitian skin effects and the breakdown of bulk-edge correspondence may have significant impact~\cite{NHSkin1}. Furthermore, possible extensions of the connection between NHTPs and DQPTs to systems in other symmetry classes, higher spatial dimensions and with many-body interactions certainly deserve further explorations.

\section*{Acknowledgement}
L.Z. is supported by the National Natural Science Foundation of China (Grant No.~11905211), the China Postdoctoral Science Foundation (Grant No.~2019M662444), the Fundamental Research Funds for the Central Universities (Grant No.~841912009), the Young Talents Project at Ocean University of China (Grant No.~861801013196), and the Applied Research Project of Postdoctoral Fellows in Qingdao (Grant No.~861905040009).

\appendix

\section{Symmetry of the geometric phase}\label{sec:AppA}
In this appendix, we analyze the symmetry of the geometric phase and
its effect on the calculation of the dynamical topological order parameter~(DTOP) for the three models considered
in this work. Due to their chiral symmetries, the Hamiltonians of
the three models in Sec.~\ref{sec:Model} share the common formalism
\begin{equation}
H(k)=d_{a}(k)\sigma_{a}+d_{b}(k)\sigma_{b},\label{eq:Hkd}
\end{equation}
where $a,b=x,y,z$ and $a\neq b$. It can be equivalently written
as
\begin{equation}
H(k)=E(k){\bf n}(k)\cdot\boldsymbol{\sigma},\label{eq:Hkn}
\end{equation}
where 
\begin{equation}
E(k)=\sqrt{d_{a}^{2}(k)+d_{b}^{2}(k)},\label{eq:Ekd}
\end{equation}
\begin{equation}
{\bf n}(k)=[n_{a}(k),n_{b}(k)]=\left[\frac{d_{a}(k)}{E(k)},\frac{d_{b}(k)}{E(k)}\right],\label{eq:nkd}
\end{equation}
and $\boldsymbol{\sigma}=(\sigma_{a},\sigma_{b})$. It is clear that
${\bf n}(k)$ is a unit vector with ${\bf n}(k)\cdot{\bf n}(k)=1$. 

According to Eqs.~(\ref{eq:Gkt}) and (\ref{eq:TP}), the total phase
of the return amplitude reads
\begin{equation}
\Phi(k,t)=-i\ln\left\{ \frac{\cos[E(k)t]}{|\cos[E(k)t]|}\right\} .\label{eq:Phikt2}
\end{equation}
It is clear that $\Phi(-k,t)=\Phi(k,t)$ once $E(-k)=\pm E(k)$. This
is clearly the case for the models considered in Subsecs.~\ref{subsec:LKC}
and \ref{subsec:NNN-LKC} according to the expressions of their bulk
spectrum $E_{\pm}(k)$. Instead, for the model studied in Subsec.~\ref{subsec:NRSSH}, the total phase does not have the parity (inversion)
symmetry.

To obtain the dynamical phase, we introduce the biorthogonal representation
of non-Hermitian systems. In this representation, the right
and left eigenvectors $\{|\psi_{s}(k)\rangle|s=\pm\}$ and $\{|\tilde{\psi}_{s}(k)\rangle|s=\pm\}$
of $H(k)$ satisfy the eigenvalue equations
\begin{equation}
H(k)|\psi_{s}(k)\rangle=E_{s}(k)|\psi_{s}(k)\rangle\label{eq:EigR}
\end{equation}
and
\begin{equation}
H^{\dagger}(k)|\tilde{\psi}_{s}(k)\rangle=E_{s}^{*}(k)|\tilde{\psi}_{s}(k)\rangle.\label{eq:EigL}
\end{equation}
The Hamiltonian $H(k)$ can also be expressed in this representation
as
\begin{equation}
H(k)=\sum_{s=\pm}E_{s}(k)|\psi_{s}(k)\rangle\langle\tilde{\psi}_{s}(k)|.\label{eq:HkBio}
\end{equation}
The time-evolution operators in the spaces of right and left eigenvectors
are 
\begin{equation}
U(k,t)=\sum_{s=\pm}e^{-iE_{s}(k)t}|\psi_{s}(k)\rangle\langle\tilde{\psi}_{s}(k)|\label{eq:UktR}
\end{equation}
and
\begin{equation}
\tilde{U}(k,t)=\sum_{s=\pm}e^{-iE_{s}(k)t}|\tilde{\psi}_{s}(k)\rangle\langle\psi_{s}(k)|,\label{eq:UktL}
\end{equation}
respectively. 

According to the definition of dynamical phase $\Phi_{{\rm D}}(k,t)$
in Eq.~(\ref{eq:DP}) of the main text, we have
\begin{equation}
\Phi_{{\rm D}}(k,t)=-\int_{0}^{t}dt'{\rm Re}\left\{ \frac{{\rm Tr}[\tilde{U}^{\dagger}(k,t')U(k,t')H(k)]}{{\rm Tr}[\tilde{U}^{\dagger}(k,t')U(k,t')]}\right\} .\label{eq:DP2}
\end{equation}
With Eqs.~(\ref{eq:UktR}) and (\ref{eq:UktL}), we can recast $\Phi_{{\rm D}}(k,t)$
into a more explicit form. The denominator of the integrand tends
out to be
\begin{equation}
{\rm Tr}[\tilde{U}^{\dagger}(k,t')U(k,t')]=2\cosh\{2{\rm Im}[E(k)]t\}.\label{eq:DENO}
\end{equation}
Furthermore, the numerator in the integrand of Eq.~(\ref{eq:DP2})
yields
\begin{equation}
{\rm Tr}[\tilde{U}^{\dagger}(k,t')U(k,t')H(k)]=2E(k)\sinh\{2{\rm Im}[E(k)]t\}.\label{eq:NUME}
\end{equation}
Putting together, we find the dynamical phase to be
\begin{alignat}{1}
\Phi_{{\rm D}}(k,t)= & -\int_{0}^{t}dt'{\rm Re}[E(k)]\tanh\{2{\rm Im}[E(k)]t\}\nonumber \\
= & -{\rm Re}[E(k)]\frac{\ln[\cosh\{2{\rm Im}[E(k)]t\}]}{2{\rm Im}[E(k)]}.\label{eq:DP3}
\end{alignat}
Referring to the main text, we see that as $E(-k)=E(k)$ for the LKC
and NNN LKC models, we also have $\Phi_{{\rm D}}(-k,t)=\Phi_{{\rm D}}(k,t)$
for these two models. Therefore, we conclude that the geometric phases
$\Phi_{{\rm G}}(k,t)=\Phi(k,t)-\Phi_{{\rm D}}(k,t)$ of the LKC and
NNN LKC models in the main text both possess the inversion symmetry,
i.e., $\Phi_{{\rm G}}(-k,t)=\Phi_{{\rm G}}(k,t)$. This allows us
to confine the range of integration to half of the first BZ, e.g.,
$k\in[0,\pi]$ for the calculation of dynamical topological order
parameters in Eq.~(\ref{eq:DTOP}) for these two models. Comparatively,
for the NRSSH model studied in Subsec.~\ref{subsec:NRSSH} of the
main text, we have $\Phi_{{\rm G}}(-k,t)\neq\Phi_{{\rm G}}(k,t)$
since $E(-k)\neq E(k)$, and the whole first BZ $k\in[-\pi,\pi]$
should be employed in the calculation of its dynamical topological
order parameter. Experimentally, information about the geometric phase may be 
directly obtained by measuring the complex spectrum dispersion $E(k)$ of the 
system~\cite{NVExp0}.

\section{DTOP of the models}\label{sec:AppB}
In this appendix, we present numerical results for the dynamical topological order parameter (DTOP) of the three models investigated in Sec.~\ref{sec:Model} of the main text. Since the geometric phase does not show any winding behaviors when there are no DQPTs, we will only consider the DTOP of the cases in which DQPTs are observed in the rate function of return probability.

\begin{figure}
	\begin{centering}
		\includegraphics[scale=0.49]{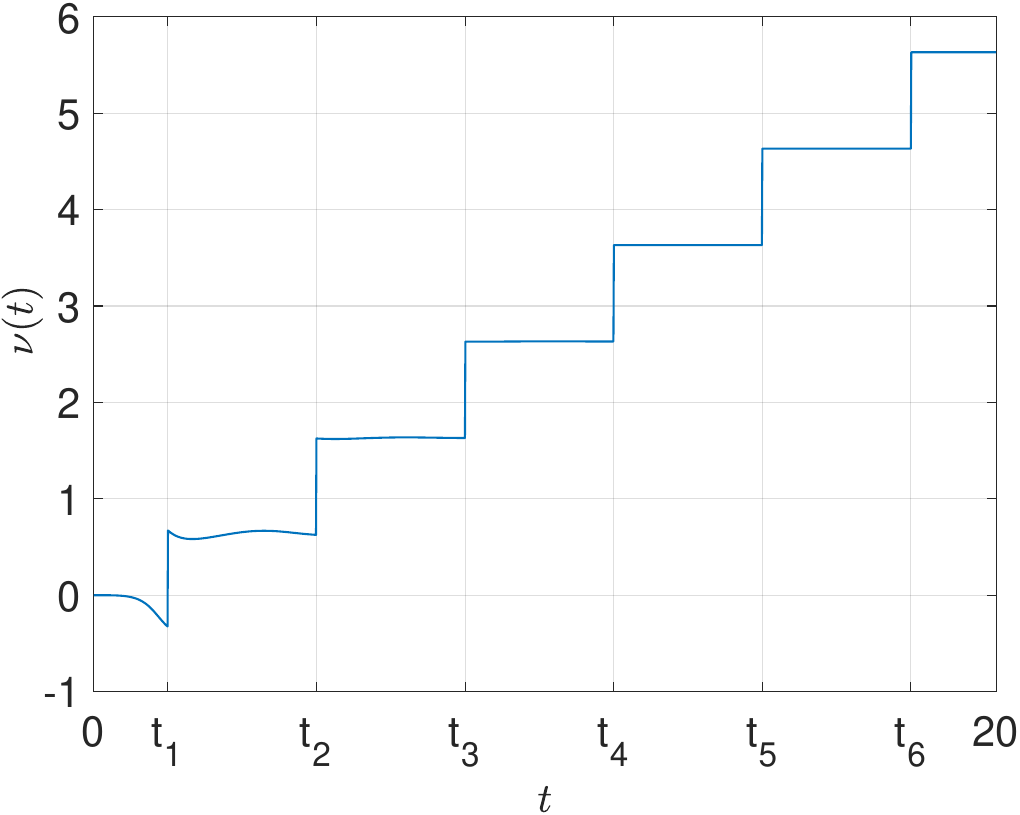}
		\par\end{centering}
	\caption{DTOP $\nu(t)$ of the LKC (blue solid line). System parameters are $J=\Delta=1$, $u=0$
		and $v=0.3$, in which the winding number $w=1$~(topological phase),
		and DQPTs are observed when the values of $\nu(t)$ possess quantized jumps. The critical times $t_{n}$~($n=1,...,6$) are obtained
		from Eq.~(\ref{eq:tnLKC}) in the main text.\label{fig:LKCDTOP}}
\end{figure}

For the LKC model defined in Subsec.~\ref{subsec:LKC}, we present the DTOP $\nu(t)$ versus time $t$ in Fig.~\ref{fig:LKCDTOP}. The system parameters are the same as those used in Figs.~\ref{fig:LKCDQPTs}(a) and \ref{fig:LKCDQPTs}(c), and the DTOP is calculated by Eq.~(\ref{eq:DTOP}) of the main text. We observe that every time when the evolution of the system passes through a critical time, the value of DTOP shows a quantized jump $|\Delta\nu(t)|=1$, which signifies the appearance of a DQPT. Meanwhile, we also notice that $\nu(t)$ may not take quantized values between certain pairs of critical times~(e.g., for $t\in(t_1,t_2)$), which might be due to our choice of reduced BZ $k\in[0,\pi]$ in the calculation of $\nu(t)$. Nevertheless, the quantized jump of DTOP across each critical time already provides us with essential information about the drastic topological change of the system when undergoing a DQPT.

\begin{figure}
	\begin{centering}
		\includegraphics[scale=0.49]{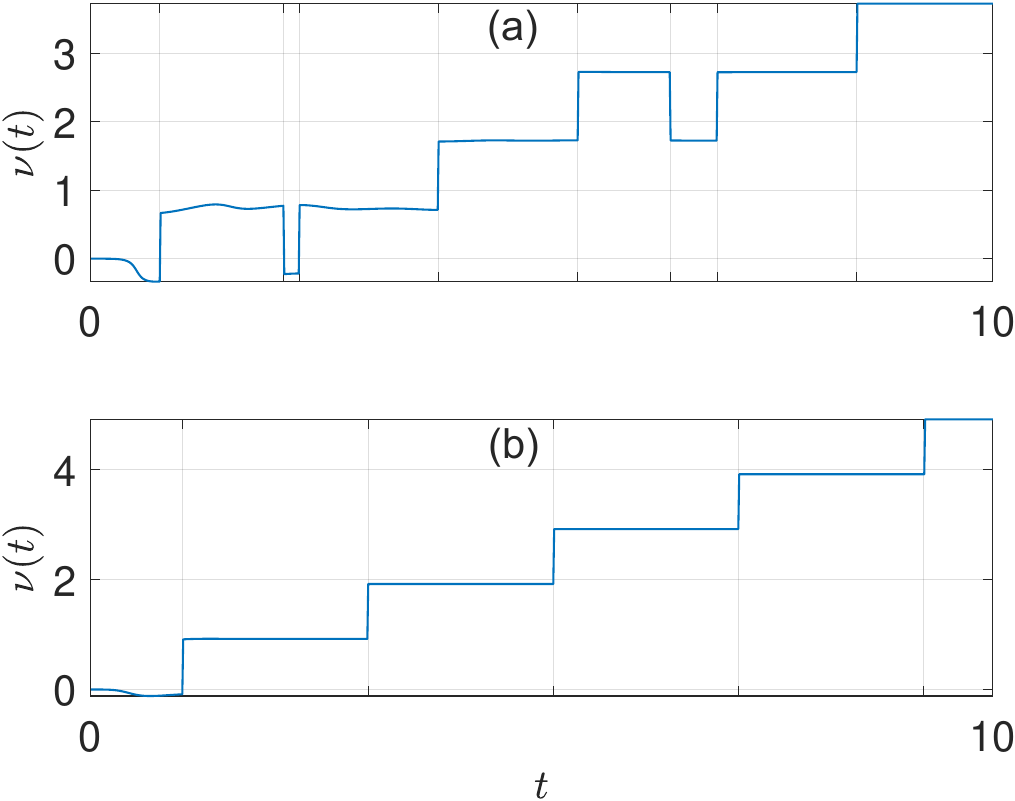}
		\par\end{centering}
	\caption{DTOP $\nu(t)$ of the NNN LKC
		model versus time. The system parameters for the post-quench Hamiltonian are $J_{1}=\Delta_{1}=1$,
		$J_{2}=\Delta_{2}=1.5$, $u=0.5$, $v=0.4$ and $1.4$ in
		panels (a) and (b), respectively. The winding numbers are
		$w=2$ and $1$ for the cases in panels (a) and (b).
		DQPTs are observed when $\nu(t)$ possesses quantized jumps in panels (a) and (b).
		In panel (a), the ticks along the horizontal axis denote the critical
		times $t_{1}(k_{c}^{+})$, $t_{1}(k_{c}^{-})$, $t_{2}(k_{c}^{+})$,
		$t_{3}(k_{c}^{+})$, $t_{4}(k_{c}^{+})$, $t_{2}(k_{c}^{-})$, $t_{5}(k_{c}^{+})$,
		$t_{6}(k_{c}^{+})$ from left to right, whose explicit values are
		obtained from Eq.~(\ref{eq:tnLKC-NNN}).
		In panel (b), the ticks along the horizontal axis are the critical
		times $t_{n}(k_{c}^{+})$ for $n=1,...,5$ from left to right.\label{fig:LKCNNNDTOP}}
\end{figure}

For the NNN LKC model defined in Subsec.~\ref{subsec:NNN-LKC}, we show the DTOP $\nu(t)$ versus time $t$ in Fig.~\ref{fig:LKCNNNDTOP}. The system parameters are chosen to be the same as those used in Figs.~\ref{fig:NNN-LKCDQPTs}(a,d) and \ref{fig:NNN-LKCDQPTs}(b,e), and the DTOP is calculated by Eq.~(\ref{eq:DTOP}) of the main text. We also notice that the value of DTOP shows a quantized jump $|\Delta\nu(t)|=1$ every time when the system evolves through a critical time, which indicates the existence of a DQPT. The applicability of DTOP to the LKC and NNN LKC models also highlights its universality in characterizing the DQPTs of 1D non-Hermitian systems with chiral symmetry.

\begin{figure}
	\begin{centering}
		\includegraphics[scale=0.49]{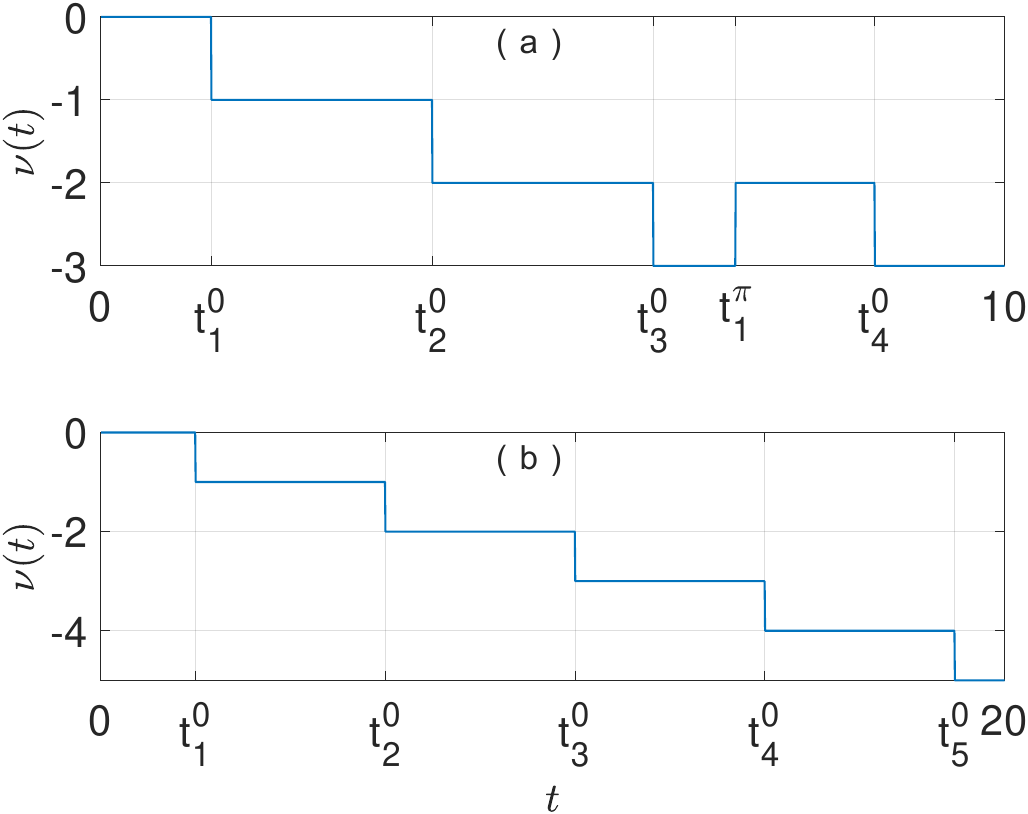}
		\par\end{centering}
	\caption{DTOP $\nu(t)$ of the NRSSH model versus time $t$.
		The system parameters for the post-quench Hamiltonian are $J_{1}=0.5$,
		$(J_{2},\gamma)=(0.8,0.2)$ and $(0.4,0.5)$ in panels
		(a) and (b), respectively. The winding numbers are $w=1$
		and $1/2$ for the cases in panels (a) and (b). DQPTs are
		observed when the values of $\nu(t)$ possess quantized jumps in panels (a) and (b). In both panels, the ticks along the horizontal axis denote the critical
		times from left to right, whose explicit values are obtained from
		Eqs.~(\ref{eq:kcNRSSH})-(\ref{eq:tpNRSSH}).\label{fig:NRSSHDTOP}}
\end{figure}

For the NRSSH model defined in Subsec.~\ref{subsec:NRSSH}, we present the DTOP $\nu(t)$ versus time $t$ in Fig.~\ref{fig:NRSSHDTOP}. The system parameters are set as the same as those used in Figs.~\ref{fig:NRSSHDQPTs}(a,d) and \ref{fig:NNN-LKCDQPTs}(b,e), and the DTOP is calculated by Eq.~(\ref{eq:DTOP}) of the main text. We find again the quantized jump of DTOP every time when the system evolves across a critical time, implying the existence of a DQPT. Furthermore, the values of DTOP remain quantized between any pair of the critical times, which is expected as the whole Brillouin zone $k\in[-\pi,\pi]$ is used in the calculation of $\nu(t)$. Besides, we also notice that the value of DTOP changes monotically in time when the post-quench system has a half-quantized winding number $w=1/2$, which is consistent with the connection bewteen the exceptional non-Hermitian topology and DQPTs as first observed in Ref.~\cite{ZhouDQPT1}.

\newpage{}

\end{document}